\documentclass[submission,copyright,creativecommons]{eptcs}
\providecommand{\event}{Seventh Workshop on Model-Based Testing 2012} 
\usepackage[pdftex]{color} 

\usepackage{listings}
\usepackage{todonotes}  

\usepackage{amssymb,amsmath} 

\newcommand{\qed}{\begin{flushright}$\square$\end{flushright}}
\newtheorem{theorem}{Theorem}[section]
\newtheorem{lemma}[theorem]{Lemma}

\title{Exact Gap Computation for Code Coverage Metrics in ISO-C}
\author{Dirk Richter
\institute{Martin-Luther-University of Halle-Wittenberg, Germany}
\email{richterd@informatik.uni-halle.de}
\and
Christian Berg
\institute{Martin-Luther-University of Halle-Wittenberg, Germany}
\email{christian.berg@student.uni-halle.de}
}
\def\titlerunning{Gap Computation for Code Coverage Metrics in ISO-C}
\def\authorrunning{Dirk Richter, Christian Berg}
\begin{document}
\maketitle

\begin{abstract}
Test generation and test data selection are difficult tasks for 
model based testing.
Tests for a program can be meld to a test suite.
A lot of research is done to quantify the quality and improve a test suite.
Code coverage metrics estimate the quality of a test suite. This quality is fine,
if the code coverage value is high or 100\%.
Unfortunately it might be impossible to achieve 100\% code coverage because of dead code for example.
There is a gap between the feasible and theoretical maximal possible code coverage value.
Our review of the research indicates, none of current research 
is
concerned with
exact gap computation.
This paper presents a framework to compute such gaps exactly in an ISO-C compatible semantic and
similar languages. We describe an efficient approximation of the gap in all the other cases.
Thus, a tester can decide if 
more tests might be able or
necessary to achieve better coverage. 
\end{abstract}

\section{Introduction}
Tests are used in model based testing to identify software defects.
High quality test generation and test data selection can be difficult tasks when
the test has to satisfy a lot of requirements or cannot be created automatically because of the undecidability of the
halting problem in Turing powerful languages.
Given requirements for a test suite (set of tests) are
functional or non-functional (e.g. execution times, runtime, usage of memory, correctness
or a minimum value of a code coverage metric). 
Code coverage metrics quantify the quality of a test suite rather imprecisely and guide testers only.
There is a gap between the feasible 
and theoretical maximal possible code coverage value.
Sometimes demanded requirements are unsatisfiable because of gaps.
Unnecessary additional tests will be computed while not all requirements are satisfied.
This 
enlarges the test suite and introduces redundancy.
Fortunately these problems (caused by metric imprecisions) can be solved by computing these gaps, which is
not possible for Turing powerful languages in general.
Therefore this paper presents suitable models in a new C-like syntax. These models allow to use an ISO-C compatible semantic.
In this paper we show how to compute such gaps exactly for these models using
formal verification techniques resp. software model checking ideas.
The paper is organized as follows: at first we clarify basics and used notations; we then present
our framework, apply it to some common coverage metrics and illustrate this on some examples.
Finally we discuss related work and present a summary and conclusions.

\vspace{-0.5em}
\section{Basics}
\label{sect:basics}

\subsection{Code Coverage Metrics $\gamma$}

Let $T_{P}=2^{tests}$ be the set of all possible sets of tests and $P$ a program written in a common programming language such as C, C++ or Java.
Each $t=\{\alpha_{1},\alpha_{2},... \} \in T_{P}$ is a test suite with tests $\alpha_{i}$ for program $P$.
The function $\gamma^{P}:T_{P} \rightarrow [0,1]$ is a code coverage metric, if $\gamma^{P}$ is monotonically increasing.
The program $P$ can be omitted, if it is well-defined by the context.


In this paper some common code coverage metrics
for functions, statements, decisions, branches and conditions will be considered as examples. Other ones
(e.g. linear code sequence and jump coverage, jj-path coverage, path coverage, entry/exit coverage or loop coverage)
can be adapted in a similar way.

The function coverage metric $\gamma_{f}^{P}(t):=|func(t)|/|func(P)|$ is the ratio of
functions $func(t)$ that has been called in the test suite $t$,
to all functions $func(P)$ in $P$ \cite{myer}.

The statement coverage metric $\gamma_{s}^{P}(t):=|stats(t)|/|stats(P)|$ is the ratio of statements
$stats(t)$ that has been executed in the test suite $t$, to all statements $stats(P)$ in $P$ \cite{myer}.
To distinguish the same statement $s$ on different program points $l_{1}$ and $l_{2}$, we
annotate each statement $s$ with unique labels $l_{1}$ and $l_{2}$ from program $P$, so that $l_{1}:s \in stats(P)$ and $l_{2}:s \in stats(P)$.

\lstset{language=c}
\begin{figure}[ht!]
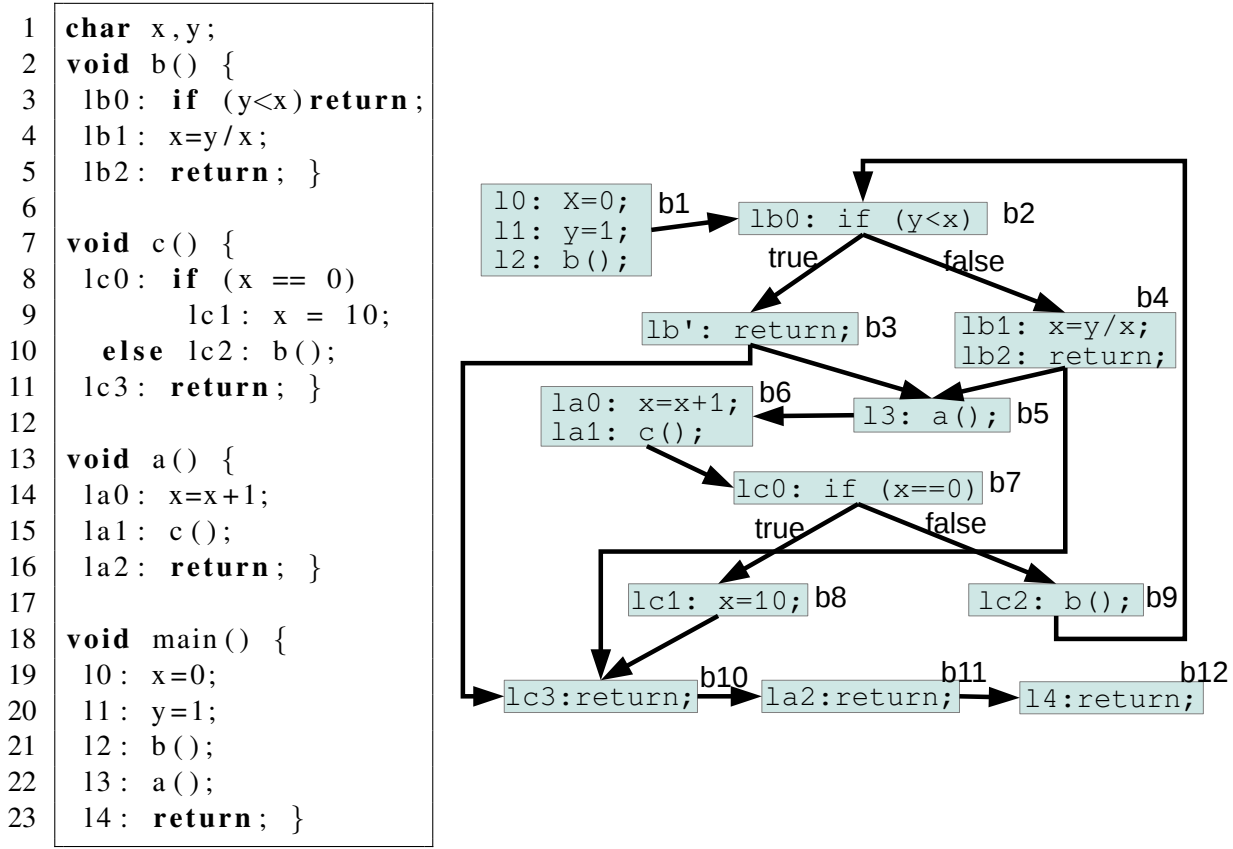

\begin{minipage}[ht!]{0.3\textwidth}
\begin{lstlisting}[frame=single,numbers=left]
char x,y;
void b() {
 lb0: if (y<x)return;
 lb1: x=y/x;
 lb2: return; }

void c() {
 lc0: if (x == 0)
       lc1: x = 10;
  else lc2: b();
 lc3: return; }

void a() {
 la0: x=x+1;
 la1: c();
 la2: return; }

void main() {
 l0: x=0;
 l1: y=1;
 l2: b();
 l3: a();
 l4: return; }
\end{lstlisting}
\end{minipage}~
\begin{minipage}[ht!]{0.7\textwidth}
\pgfdeclareimage[interpolate=true,width=11cm]{pic1}{cfglisting1}
\pgfuseimage{pic1}
\end{minipage}
\caption{SPDS example $P_{1}$ in ISO-C syntax and corresponding BBICFG}
\label{bsp}
\end{figure}

Let $blocks(P)$ be all basic blocks \cite{basicblock} in $P$. Every program point has a surrounding basic block.
The basic block inter-procedural control flow graph $BBICFG_{P}=(blocks(P),edges(P))$ (see Fig. \ref{bsp}) consists of the basic blocks $blocks(P)$ as nodes and edges
$edges(P) \subseteq blocks(P)^{2}$, where $(b_{1},b_{2}) \in edges(P)$ iff there is an execution path of length $1$ from the end of block $b_{1}$
to the entry of block $b_{2}$ (execution of the last statement of block $b_{1}$).
The decision coverage metric $\gamma_{d}^{P}(t):=|edges(t)|/|edges(P)|$ is the ratio
of executed edges $edges(t)$ of the control flow graph $BBICFG_{P}$ for $t$, to all edges $edges(P)$ in $P$.

The branch coverage metric $\gamma_{b}^{P}(t):=|blocks(t)|/|blocks(P)|$ is the ratio of basic code blocks $blocks(t)$
executed during test suite $t$, to 
all basic blocks $blocks(P)$ in $P$ \cite{branch}.
Even if all basic blocks are covered by test suite $t$ and $\gamma_{b}^{P}(t)=1$, there can be uncovered branching edges in the
basic block inter-procedural control flow graph $BBICFG_{P}$. Thus $\gamma_{d}^{P}(t)<1$ is possible in this case.

Let $bExpr(l)$ be the set of all Boolean sub-expressions on label $l$ of program $P$ and
\begin{equation}
BExpr(P):=\{ (l,bExpr(l)) ~\bullet~ l \in labels(P) \}.
\end{equation}
The condition or predicate coverage metric $\gamma_{c}^{P}(t):=|exval(t,P)|/(2\cdot |BExpr(P)|)$
is the ratio of evaluations of boolean sub-expressions $exval(t,P) \subseteq BExpr(P) \times \{true,false\}$
of the test suite $t$, to all evaluations of boolean sub-expressions in $P$ \cite{myer}.
The relation $exval(t,P)$ describes the evaluations of sub-expressions $e$ on label $l$ 
under test suite $t$, such that 
$((l,e),true) \in exval(t,P)$ iff there is a test $\alpha \in t$ where $e$ can be evaluated to $true$ on label $l$ under test $\alpha$.
When boolean operations are not short circuited, condition coverage does not necessarily imply decision coverage. 

\subsection{Code Coverage Metric Gap $\delta$}

Let $\gamma:T_{P} \rightarrow [0,1]$ be a code coverage metric for a program $P$.
The code coverage metric gap 
$\delta_{\gamma}(P) \in [0,1]$
is the smallest difference between the coverage ratio of a test suite $t \in T_{P}$ and the theoretical maximal value 1:
\begin{equation}
\delta_{\gamma}(P):= \inf_{t \in T_{P}} (1-\gamma(t)).
\end{equation}
Let $\delta_{x}(P)$ denote $\delta_{\gamma_{x}}(P)$, where $x \in \{c,d,s,f,b\}$.
Obviously dead code can cause $\delta_{\gamma}(P)>0$.
If some evaluations of boolean (sub-)expressions cannot be realized,
$\delta_{\gamma}(P)>0$ is possible without dead code (e.g. condition or decision coverage).
In Turing powerful programming languages the gap $\delta_{\gamma}(P)$
can not be computed in general, because the halting problem is undecidable.
In this case the gap $\delta_{\gamma}(P)$ can be approximated only. We show how to compute the exact gap $\delta_{\gamma}(P)$ for
an ISO-C compatible semantic by adequate modeling.

\subsection{Suitable Models}
\label{subsec:suit}

A more expressive model describing the
program behaviour allows for 
a more accurate
approximation of the gap $\delta_{\gamma}(P)$.
If the model is not Turing powerful and the model behaviour is equivalent to the program behaviour, the gap
$\delta_{\gamma}(P)$ can be exactly computed. 
Therefore we defined an ISO-C compatible semantic using pushdown systems (PDS) \cite{kps09rai}.
The split of the ISO-C language definition into platform-independent semantics and platform-specific semantics has
a serious implication for deciding the halting problem of ISO-C programs: whether a C program halts or not
depends on the platform-specific semantic. Thus, even though the halting problem
for a C program is decidable for a platform-specific semantic, the halting
property can become undefined if no specific platform is assumed \cite{kps09rai}.
Now we present an extension of the PDS used in \cite{kps09rai} to symbolic pushdown systems (SPDS) using an ISO-C like syntax.
SPDS use a more compact representation and define the PDS configurations and transitions symbolically.

A SPDS is a tuple $S=(vgbl,func)$, where $vgbl$
is a finite set of variables (global variables in ISO-C) and
$func$ is a set of functions (pairwise different names) with an initial function $main \in func$.
Each variable $v$ has an integer type\footnote{The Boolean constants $false$ and $true$ are represented
via $0$ and $\neq 0$ like ISO-C.}
$bits(v) \in \mathbb{N}_{\geq 1}$ and a fixed length $len(v) \in \mathbb{N}_{\geq 1}$.
Every variable $v$ is an array. 
A function is a tuple $(f,param,vlcl,stats$), where
$(f,param)$ is a function signature with a unique function name $f$ and a finite list of parameter variables $param$.
The set $vlcl$ is a finite set of variables (local variables in ISO-C), such that $param \subseteq vlcl$.
The body of $f$ is a finite list of statements $stats$.
Each statement $l:s \in stats$ has a unique label $l \in labels(f)$,
and $fst(f) \in labels(f)$ is the label of the first statement in the list $stats$. 
We use 
 $vgbl=vgbl(S), func=func(S), param=param(f), vlcl=vlcl(f)$ and $stats=stats(f)$ respectively,
 if $S$ or $f$ are well-defined by the context.
Let denote $func(l)$ the function $f$ for which $l \in labels(f)$. Further let be
\begin{equation}
vars:=vgbl\cup \bigcup_{f\in func} vlcl(f)~,~~ stats(S):=\bigcup_{f\in func} stats(f) \textnormal{~ and ~} labels(S):=\bigcup_{f\in func} labels(f).
\end{equation}
Similar to ISO-C the SPDS variables are used to build expression $Expr$ using constants and operators.
The priority and associativity are the same as in ISO-C.

An expression $e \in Expr$ can be strictly evaluated to an integer number $[\![e]\!]_{g}^{c_{f}} \in \mathbb{Z} \cup \{\bot\}$
using valuation functions for global and local variables $g:vgbl \times \mathbb{Z} \rightarrow \mathbb{Z}$ and
$c_{f}:vlcl(f) \times \mathbb{Z} \rightarrow \mathbb{Z}$.
The symbol $\bot$ denotes arithmetic exception (e.g. division-by-zero or index-out-of-bounds). 
The functions $g(v,i)$ and $c_{f}(v,i)$ return the current value of variable $v$ at index $i$ (value of $v[i]$).
The evaluation functions $g$ and $c_{f}$ can be omitted, if they are well-defined by the context.
A variable usage $v[i]$ of variable $v \in vars$
with index $i \in \mathbb{Z}$ 
is evaluated as
\begin{equation}
[\![~v[i]~]\!]_{g}^{c_{f}}:=\begin{cases} g(v,i) & v\in vgbl \wedge 0 \leq i < len(v) \\
                                c(v,i) & v \in vlcl(f) \wedge 0 \leq i < len(v) \\
                                \bot & otherwise. \end{cases}
\end{equation}
For $e \in Expr$ and a statement $l:s \in stats(f)$, $s$ has one of the following forms:
\begin{itemize}
\item $v[e_{1}] = e_{2};$ ~~~~~~~corresponds to writing the value $[\![e_{2}]\!]$ into the variable $v$ at index $[\![e_{1}]\!]$.
\item $f(v_1,\ldots,v_n);$ ~~~corresponds to a function call (call by value), iff $(f,param)$ is a signature, where \\
\hspace*{2.4cm}$param=[p_{1}, p_{2}, \dots, p_{n}]$, $v_{i} \in vars$
and $bits(v_{i})\leq bits(p_{i})$ for all $1\leq i \leq n$.
\item $return; $ ~~~~~~~~~~~~~corresponds to a function return.
\item $if~(e)~goto~l';$ ~~corresponds to a conditional
  jump\footnote{intra-procedural} to label $l' \in labels(f)$.
\end{itemize}
The exception of a dynamic type mismatch occurs for "$v[e_{1}] = e_{2};$" and the system terminates,
if $[\![e_{2}]\!]=\bot$ or the type $bits(v)$ is too small to store the
value $[\![e_{2}]\!]$ or $[\![e_{1}]\!] \notin \{0,1,\dots,len(v)\}$. 
We denote $v=e$ for $v[0]=e$ and $v$ for usages of $v[0]$ to emulate syntactically non-array variables.
The system terminates on statement "$if~(e)~goto~l';$" too, if $[\![e]\!]=\bot$.
The predefined function $rand(e)$ returns a random number between $0$ and $[\![e]\!]$ for $e \in Expr$, whereby $rand(\bot)=\bot$.
Further ISO-C statements and variations for other languages can be mapped to these
basic statements in the modeling phase. 
All variables (global and local) are uninitialized and have initially a random value.
A test $\alpha$ for $S$ is a subset of global variables with predefined values for label $fst(main)$. 
A configuration
$s=(g,[(l_{n},c_{n}), (l_{n-1},c_{n-1}),\dots,(l_{1},c_{1})])$  of $S$ represents a state of the underlying Kripke structure
 with the current execution label $l_{n} \in labels$, the valuation
$g:vgbl \times \mathbb{Z} \rightarrow \mathbb{Z}$ of global variables and the stack content.
The stack content consist of a list of function calls with
current execution labels $l_{i} \in labels(S)$ and
valuations for local variables $c_{i}:vlcl(func(l_{i})) \times \mathbb{Z} \rightarrow \mathbb{Z}$.
The head of $s$ is $head(s)=(g,(l_{n},c_{n}))$. The set of all possible configurations is $conf(S)$.
A $run$ of $S$ is a sequence of consecutive configurations beginning with an initial configuration 
$(g_{init},[fst(main),c_{init}]) \in conf(S)$.
SPDS are (like PDS) not Turing powerful and can be used to model the behaviour of (embedded) ISO-C programs. 
There is no restriction on the recursion depth. 

\section{Exact Gap Computation Framework}
\label{sect:exact}

Let $\gamma:T_{P} \rightarrow [0,1]$ be a code coverage metric for a program $P$.
Our framework to compute the gap $\delta_{\gamma}(P)$ consists of the following steps:
\begin{enumerate}
    \item If necessary, create a SPDS model $S$ with ISO-C compatible semantic for program $P$.\vspace{-0.1cm}
    \item Modify the model $S$ to a SPDS model $S'$ to enable gap analysis for the code coverage metric $\gamma$.\vspace{-0.1cm}
    \item Compute exact variable ranges for some new variables in $S'$.\vspace{-0.1cm}
    \item Conclude the exact size of the gap $\delta_{\gamma}(S)$ in $S$ for the code coverage metric $\gamma$.\vspace{-0.1cm}
    \item Conclude the size of the gap $\delta_{\gamma}(P)$ in $P$.\vspace{-0.1cm}
\end{enumerate}

\subsection{SPDS Modeling (step 1)}
\label{isomodel}

If the given program $P$ is not written in ISO-C (e.g. Java) or $P$ has another
platform-specific semantic, we create a SPDS model $S$ for $P$ by abstraction. Otherwise the behaviour of $S$
is the same of $P$ by mapping all the ISO-C statements to the basic SPDS-statements of section \ref{subsec:suit} using abbreviations (described
in this section). 
Java can be handled using the tool JMoped \cite{jmopedreport}. Often other languages and corresponding statements can be mapped
to the basic SPDS-statements in a similar fashion. 
For simplification 
we present some common mappings, which are abbreviations for previously defined basic SPDS-statements.
We sketch the ideas only, because of limited space. Fig. \ref{bsp} shows the SPDS example $P_{1}$ in ISO-C syntax, where "$\texttt{char~x}$"
in line $1$ is an abbreviation for
"$\texttt{int~x(8)[1]}$" to declare an integer array of type $bits(x)=8$ and $len(x)=1$.
\\
\textbf{Omitted Returns and Labels:}
If there is no return statement at the end of a function body, its existence is assumed during interpretation
of the symbolical description of $S$. The same holds for statements without labels, such that each statement
in the SPDS has a unique label after interpretation. 
\\
\textbf{Parameter Expressions:}
Basic SPDS-statements allow variables to be passed as parameters in function calls. We can simulate to pass expressions
by temporary local variables. Let "$l:f(e_{1},e_{2},\dots,e_{n});$" be a function call with expressions $e_{i} \in Expr$,
where $(f,param)$ is a signature with $param=[p_{1}, p_{2}, \dots, p_{n}]$.
We introduce new 
local SPDS-Variables $pe_{i} \notin vars$ with type
$bits(pe_{i})=bits(p_{i})$ and $len(pe_{i})=1$. These variables are used to evaluate the expressions before the function call: $pe_{i}=e_{i}$.
Instead of $e_{i}$ now $pe_{i}$ is passed to $f$ using the
basic SPDS-statement $f(pe_{1},pe_{2},\dots,pe_{n})$.
The function call "$l:f(e_{1},e_{2},\dots,e_{n})$" is interpreted as
"$l:pe_{1}=e_{1}; pe_{2}=e_{2}; \dots pe_{n}=e_{n}; f(pe_{1},pe_{2},\dots,pe_{n})$".
Now only basic SPDS-statements are used.
The code coverage metrics are adapted accordingly.
For example the statements $pe_{i}=e_{i};$ are ignored for
the statement coverage metric.
\\
\textbf{Return Values:}
A function can return a value. This value can be used to set a variable "$v=f(...);$".
If a function returns an expression $e$ via "$return~e$", a new global variable $ret_{f} \notin vars$ is introduced.
The type of $ret_{f}$ equals the return type of function $f$ and $len(ret_{f})=1$.
The statement "$return~e;$" is interpreted as "$ret_{f}=e; return;$".
On the other hand the assignment "$v=f(...);$" is interpreted as "$f(...); v=ret_{f};$" to store the return value of $f$ in $v$.
\\
\textbf{Function Calls in Expressions:}
If there is a function call $f(..)$ in an expression $e$, this function is evaluated in a temporary local variable.
Boolean operations in SPDS are strict and not short circuited. 
Short circuited expressions (also increments \texttt{i++} and decrements \texttt{i--}) can be mapped to
strict expressions without side effects by several conditional statements. Thus every function call $f(..)$ in an
expression $e$ will be definitively evaluated during the evaluation of $e$. Accordingly it is safe to
do every call before the evaluation of $e$. 
Sometimes the order of this evaluation is implementation defined (as in ISO-C) and depends on the source language (e.g. \texttt{i++*i++}).
Thus
we use priority and associativity for calculating this order.
The intermediate representation of a compiler can be used, too, to achieve
a mapping to SPDS.
\\
\textbf{Unconditional Jump:}
The statement "$goto~l;$" is mapped to "$if~(1)~goto~l;$". The dead branching edge to the following statement is ignored by the decision coverage etc.
\\
\textbf{Skip Statement:} Particularly low level languages have often a "no operation" statement. We use the statement $skip$,
which does not change variable settings.
The statement "$l:skip;$" can be interpreted using the conditional branch "$l:if~(0)~goto~l;$".
If there is a global variable $v \in vars$, this can also be interpreted as "$l:v=v;$". The former needs no consideration
for the coverage metrics.
\\
\textbf{Random Numbers:}
In ISO-C the function $rand()$ returns a pseudo random value between $0$ and the constant $RAND\_MAX$, where $RAND\_MAX$ depends on the system.
We can map this behaviour using the SPDS function $rand(RAND\_MAX)$. A similar mapping for random numbers is possible in other languages.
\\
\textbf{Conditional Statements:}
Let $s1$ and $s2$ be lists of statements.
The conditional statement "$if~(e)$ $s1~else~ s2;$" is interpreted as
"$if~(e)~goto~l_{1}; s2; goto~l'; l_{1}: s1; l': skip;$", where $l_{1},l' \notin labels$.
"$if~(e)~s;$" is an abbreviation for "$if~(e)~s~else~skip;$". 
\\
\textbf{Local Variable Definitions:}
A local variable can be defined during an assignment of a basic block or a loop header.
Such local variable definitions are mapped to local variables of the surrounding
function. Renaming can be done easily if necessary. 
\\
\textbf{Loops:}
A for-loop of the form "$for~ (init;~ cond;~ inc)~ body;$" is interpreted as "$init;$ $l:$ $body;$ $inc;$ $if~(cond)$ $goto$ $l;$",
where $l \notin labels$. The $do$ and $while$ loops are interpreted in a similar way.
\\
\textbf{Modular Arithmetic and Integer Overflow:}
The ISO-C standard says that an integer overflow causes
"undefined behaviour", meaning that compilers conforming to the standard
can generate any code: from completely ignoring the overflow to aborting
the program. Our solution terminating the system is conform to the ISO-C standard.
Evaluations of expressions in SPDS are not restricted to arithmetic bounds, but dynamic type mismatches
are possible for assignments $v=e$.
In the case of modeling nonterminating modular arithmetic the modulo operator $\%$ can be used
to shrink the expression $e$ to fit the size $bits(v)$. Hence, a dynamic type mismatch does not occur.
\\
\textbf{Dynamic Memory and Pointers:}
In ISO-C a certain amount of the heap can be reserved using the function $malloc(int)$.
It returns an address on the heap. The heap is finite, because the number of addresses
is finite. This behaviour is simulated using a global array $heap$ of type $bits(heap)=8$ with length $len(heap)=m$ and a global variable $ptr$
with type $bits(ptr)=\lceil log_{2}(m) \rceil$ and $len(ptr)=1$, which points to the next free space in the heap array.
The function $malloc(int)$ can be implemented as shown in Listing \ref{malloc} with $1024$ heap elements respectively, which
needs a $10$ bit variable $ptr$ for accessing.
 A memory exceptions occurs (label $memout$),
if there is not enough memory left.
\lstset{language=C}
\begin{lstlisting}[frame=single,label=malloc,caption=Malloc as SPDS in ISO-C like syntax,
captionpos=b,numbers=left,
commentstyle=\small\ttfamily,
basicstyle=\small\ttfamily]
int heap(8)[1024];
int ptr(10);

int(10) malloc(int n(10)) {
  if (ptr >= 1024-n) goto memout;
  ptr = ptr+n;
  return ptr-n; }
\end{lstlisting}
Once reserved space can be reused, because a garbage collector and a function $free$ can be implemented in SPDS.
A pointer is a SPDS variable used as an index of the heap array and an address is just another index (returned by the address operator $\&$).
Variables placed in the heap array support the address operator in contrast to the other SPDS variables.
If putting a \textbf{local} variable of a function $f$ into the heap array, the recursion of $f$ will be bound, because of a finite maximal heap size.
Coverage metrics 
have to adapt to these additional SPDS functions, statements and variables to be able to compute the
correct gap.
\\
\textbf{Call by Reference:}
Instead of passing a variable as a function parameter, a pointer can be used to indirectly access variable values in the heap.
Thus call by reference can be simulated. Unfortunately this results in bounding the recursion, too.
\\
\textbf{Dynamic Arrays:}
Array semantic in ISO-C is defined by pointers and access to its elements is defined by pointer arithmetic.
Thus $malloc(int)$ can be used for this purpose. 

Other constructs and statements from other languages (e.g. classes, structs, objects, dynamic parameter lists, etc.)
can be mapped in a similar way.
If an arithmetic exception occurs, the SPDS ends and the corresponding ISO-C program $P$ can
have undefined behaviour according to the language specification.
$P$ can terminate, which is a complying behaviour. Therefore this behaviour is used for our modeling process. 
Other implementation defined behaviour can be modeled similarly.

\subsection{Extraction of Exact Variable Ranges (step 3)}

In step 2 a SPDS $S'$ is created for the SPDS $S$ by slightly modifying $S$ (explained in the next section).
For a PDS $B$ an automaton $Post^{*}(B)$ can be computed, which accepts all reachable configurations of $B$\cite{Sch02}.
Thus 
for the SPDS $S'$ a similar automaton $Post^{*}(S')$ can be created, because $S'$ is just a symbolical PDS.
This is a basic step in symbolic model checking using Moped \cite{moped}. We use the
$Post^{*}$ algorithm of the model checker Moped for our implementation by mapping our SPDS definitions to
the input language Remopla\footnote{e.g. we map integers to nonnegative numbers, as Remopla does not support negative integers} \cite{Kie06}.
The set of reachable heads $h(S'):=\{ (g,(l,c)) ~\bullet~ (g,[(l,c)...]) \in Post^{*}(S')\}$ is finite because of finite variable types.
Thus 
exact variable ranges can be extracted from $h(S')$.
Let $v \in vars$ and $l \in labels$,
then $range_{l}^{S'}(v):=\{ [\![v]\!]_{g}^{c} ~\bullet~ (g,(l,c)) \in h(S') \}$
is the exact variable range of $v$. The notation $S'$ can be omitted, if $S'$ is well-defined by the context.
For all values $k \in range_{l}(v)$ there is a run of $S'$, such that $[\![v]\!]=k$ on label $l$ and vice versa.

$h(S')$ and $range_{l}(v)$ can be computed symbolically out of $Post^{*}(S')$ using Ordered Binary Decision Diagrams (OBDD) operations.
The computation of $h(S')$ is a straightforward OBDD restriction operation in $Post^{*}(S')$ and results in a
characteristic function $q:\{0,1\}^{n} \rightarrow \{0,1\}$ represented as an OBDD.
The input vectors of $q$ are heads $h(S')$ encoded as finite Bit sequences. The computation of $range_{l}(v)$ uses cofactors.
 A cofactor of $q$ is
$q[x_{i}=b](x_{1}, x_{2}, \dots, x_{n}):= q(x_{1}, x_{2}, \dots,x_{i-1},b,x_{i+1} \dots, x_{n})$ \cite{kps09dirk}.
The positive cofactor is
$q[x_{i}=1]$ and the negative cofactor is $q[x_{i}=0]$.
A characteristic function $r:\{0,1\}^{m} \rightarrow \{0,1\}$ for $range_{l}(v)$ can be computed using
cofactors:
\begin{lemma}
Let $k$ be the starting index of the encoding of $v$ on label $l$ in $q$ and let $m$ be the length of the encoding.
Then
$r(y_{1},y_{2},\dots,y_{m})=1$ is valid, iff $q[x_{k}=y_{1}][x_{k+1}=y_{2}]\dots [x_{k+m-1}=y_{m}]$
is not always $0$ (not the empty OBDD).
\end{lemma}
The proof is a consequence of the definitions.
The computation of exact variable ranges is more time-consuming than model-checking the reachability
in $S'$ \cite{kps09dirk}.
Fortunately $range_{l}(v)$ can be approximated using static data flow analyses and test suites.
This is the case for focusing on efficiency or unbounded recursion depth in
combination with unbounded parallelism.
For further reference in comparisons, explanations, and proofs see \cite{kps09dirk}.

\subsection{SPDS Supplementation (step 2) and Exact Gap Inference (step 4)}

Now we show exemplary, how to apply our framework to common code coverage metrics.

\subsubsection{Function Coverage Gap $\delta_{f}(S)$}

We supplement $S$ with a new global variable $v \notin vgbl(S)$ using type $bits(v)=1$ and $len(v)=1$
without any assignment or reading usage on $v$ to ensure the existence of at least one global variable in $S'$.
By construction this variable $v$ has a random undefined value $[\![v]\!] \in \{0,1\}$ on each label $l \in label(S')$ resp. on each program point.
The exact function coverage gap $\delta_{f}(S)$ can be concluded from the exact ranges of variables in $S'$ as follows:
\begin{lemma} \label{lem1}
    \begin{equation}
    \delta_{f}(S)= 1-\frac{|\{f \in func(S) ~\bullet~ range_{fst(f)}^{S'}(v) \neq \emptyset \}| }{|func(S)|}
    \end{equation}
\end{lemma}
\begin{bew}
The new global variable $v$ does not influence the model behaviour. All variable evaluations and reachable labels in $S'$
are the same in $S$. Further it is $vgbl(S') = vgbl(S) \cup \{v\}$, $func(S')=func(S)$ and
$func(S) \neq \emptyset$ because of $main \in func(S)$.
The main observation is, that a label $l \in labels(S)$ is unreachable or dead, iff $range_{l}^{S'}(v)= \emptyset$.
Thus a function $f$ can be called, iff $range_{fst(f)}^{S'}(v) \neq \emptyset$.
Choose a test suite $t' \in T_{S}$ such that $|func(t')|$ is maximal.
With the maximality of $t'$ we have
\begin{equation}
|func(t')| \geq |\{f \in func(S) ~\bullet~ range_{fst(f)}^{S'}(v) \neq \emptyset \}|.
\end{equation}
Every function $f \in func(t')$ has a cover witness test $\alpha \in t'$, so that the label
$fst(f)$ is reachable under test $\alpha$.
Thus it is $range_{fst(f)}^{S'}(v) \neq \emptyset$.
On the other hand we obtain
\begin{equation}
|func(t')| \leq |\{f \in func(S) ~\bullet~ range_{fst(f)}^{S'}(v) \neq \emptyset \}|,
\end{equation}
because
each $f \in func(S)$ with $range_{fst(f)}^{S'}(v) \neq \emptyset$ has 
at least
one test $\alpha'$ (not necessarily $\in t'$) to cover the function $f$, which can be detected by an evaluation of $v$.
Accordingly it is
\begin{equation}
\sup_{t\in T_{S}} |func(t)| = |\{f \in func(S) ~\bullet~ range_{fst(f)}^{S'}(v) \neq \emptyset \}|,
\end{equation}
 which
is equivalent to
\begin{equation}
\inf_{t\in T_{S}} (1-\gamma_{f}(t)) = 1-\frac{|\{f \in func(S) ~\bullet~ range_{fst(f)}^{S'}(v) \neq \emptyset \}| }{|func(S)|}.
\end{equation} \qed
\end{bew}

\vspace{-0.2cm}
The exact branch coverage gap $\delta_{b}(S)$ can be computed similarly. Instead of function entry points, just the block
entry points are considered. 

\subsubsection{Statement Coverage Gap $\delta_{s}(S)$}

The SPDS $S'$ will be supplemented with a new variable $v~\notin~vgbl(S)$ and
the type $bits(v)=1$ and $len(v)=1$ similarly to the function coverage gap.
The exact statement coverage gap $\delta_{s}(S)$ can be computed:
\begin{lemma}
    \begin{equation}
    \delta_{s}(S)= 1-\frac{|\{l:s \in stats(S) ~\bullet~ range_{l}^{S'}(v) \neq \emptyset \}| }{|stats(S)|}
    \end{equation}
\end{lemma}
\begin{bew} Similar to Lemma \ref{lem1} prove
$\sup_{t\in T_{S}} |stats(t)| = |\{l:s \in stats(S) ~\bullet~ range_{l}^{S'}(v) \neq \emptyset \}|$. \qed
\end{bew}

\subsubsection{Decision Coverage Gap $\delta_{d}(S)$}

The branch coverage uses the nodes of the control flow graph $BBICFG_{S}$ and
the decision coverage uses the edges.
The execution of an edge $(b_{1},b_{2}) \in edges(S)$ in the control flow graph $BBICFG_{P}$ depends on several conditions such as
arithmetic overflow, division-by-zero or boolean expressions for conditional branches.
To compute the exact decision coverage gap, we introduce a new global variable $v_{in} \notin vars(S)$ into $S'$ with type
$bits(v_{in})=1+\lceil log_{2}(|blocks(S)|) \rceil$ and $len(v_{in})=1$. Each label $l$ belongs to a basic block $b_{l}$, which can be identified
by a unique number $n_{b_{l}} \in \mathbb{N}_{\geq 0}$. This number is assigned to the variable $v_{in}$ to
detect the past basic block for a statement.
The type $bits(v_{in})$ is big enough to store every unique identifier $n_{b_{l}}$. Each statement $"l:s" \in stats(S)$ is
modified\footnote{Additionally this can be done using the native synchronous parallelism
in SPDS without an extra label: "$l: v_{in}=n_{b_{l}},~s$".} to
"$l: v_{in}=n_{b_{l}};~ l': s$" in $S'$, where $l' \notin labels(S)$ is unique. So it is possible to determine
the past basic block on label $l$ using the exact range of the SPDS variable $v_{in} \in vgbl(S')$.
The exact decision coverage gap $\delta_{d}(S)$ can be computed:

\begin{lemma}
    \begin{equation} \label{dcg}
    \delta_{d}(S)= 1-\frac{|\{ (a,b) \in edges(S) ~\bullet~ n_{a} \in range_{fst(b)}^{S'}(v_{in}) \}| }{|edges(S)|}
    \end{equation}
\end{lemma}
\begin{bew} Let $a,b \in blocks(S)$ be basic blocks. 
By construction it is $n_{a} \in range_{fst(b)}^{S'}(v_{in})$, iff there is an execution path from the end of basic block $a$ 
directly to the first label $fst(b)$ of basic block~$b$ in $S$.
This is equivalent to the existence of a test $\alpha$, such that $(a,b) \in edges(\alpha)$.
Thus we have
\begin{equation}
\exists \alpha \in t': (a,b) \in edges(\alpha) \Leftrightarrow n_{a} \in range_{fst(b)}^{S'}(v_{in})
\end{equation}
for a
chosen $t' \in T_{S}$, where $|edges(t')|$ is maximal. Hence it is
\begin{equation}
\sup_{t\in T_{S}} |edges(t)| = |\{ (a,b) \in edges(S) ~\bullet~ n_{a} \in range_{fst(b)}^{S'}(v_{in}) \}|,
\end{equation}
 which shows
(\ref{dcg}) similar to
Lemma \ref{lem1}.\qed
\end{bew}

\vspace{-0.5cm}
\subsubsection{Condition Coverage Gap $\delta_{c}(S)$}

For the condition coverage all boolean sub-expressions (conditions $BExpr(S)$) on each label are considered.
The theoretical maximal value can be achieved, when every condition $(l,e) \in BExpr(S)$ can be $1$~($true$) and $0$ ($false$).
For each boolean sub-expression $b \in B=\bigcup_{l \in labels(S)} bExpr(l)$ we introduce
new boolean global variables $v_b \notin vars(S)$ with type $bits(v_b)=1$ and $len(v_{b})=1$ into $S'$.
Let further $bExpr(l)=\{e_1, e_2, \dots e_n\}$ be the set of all boolean sub-expressions on label $l$.
Each statement $"l:s" \in stats(S)$ with $bExpr(l)\neq \emptyset$ will be modified to
"$l: v_{e_1}=e_{1};~v_{e_2}=e_{2};~\dots v_{e_n}=e_{n};~ l': s$" in $S'$, where $l' \notin labels(S)$ is unique.
The statement $"l:s" \in stats(S)$ will be modified to
"$l: skip;~ l': s$" in $S'$, if $bExpr(l)= \emptyset$.
Hence the existence of label $l' \in labels(S')$ is guaranteed.
Thus the exact condition coverage gap $\delta_{c}(S)$ can be computed:

\begin{lemma}
    \begin{equation} \label{ccg}
    \delta_{c}(S)= 1-\frac{ {\sum \atop {l \in labels(S) \atop e \in bExpr(l)}} |range_{l'}^{S'}(v_{e})|}{2 \cdot |BExpr(S)|}
    \end{equation}
\end{lemma}
\begin{bew} Choose a $t' \in T_{S}$ such that $|exval(t',S)|$ is maximal.
Then it is
\begin{eqnarray}
    && ((l,e),b) \in exval(t',S) \\ 
    &\Leftrightarrow&  \textnormal{expression $e$ can be evaluated to $b \in \{0,1\}$ on label $l$ in $S$} \\
    &\Leftrightarrow& b \in range_{l'}^{S'}(v_{e}).
\end{eqnarray}
This proves (\ref{ccg}) 
similar to Lemma \ref{lem1}, because of
\begin{equation}
\sup_{t\in T_{S}} |exval(t,S)| = \sum_{l \in labels(S) \atop e \in bExpr(l)} |range_{l'}^{S'}(v_{e})|.
\end{equation}\qed
\end{bew}

\vspace{-0.6cm}
\subsection{Conclusion for $\delta_{\gamma}(P)$ based on $\delta_{\gamma}(S)$ (step 5)}

The computed gap $\delta_{\gamma}(S)$ is exact ($\delta_{\gamma}(S) = \delta_{\gamma}(P)$),
if the behaviour of $S$ is equivalent to the behaviour of $P$.
Hence step 1 does not abstract nor simplify the program behaviour. 
This is the case for our ISO-C compatible semantic \cite{kps09rai} on C programs.
If $S$ abstracts from the behaviour of $P$, $\delta_{\gamma}(S)$ is only an approximation for $\delta_{\gamma}(P)$.
The approximation degree depends on the degree of this abstraction.

\vspace{-0.7em}
\section{Gap Approximation using $\delta_{\gamma}^{-}$ and $\delta_{\gamma}^{+}$}
\label{sect:approx}

The gap can be approximated by
abstracting the program $P$ to a simpler behavior of SPDS $S$ as shown above.
On the other hand the exact variable ranges $range_{l}(v)$
can be approximated, too. This is a more practical approach
particularly for huge software systems.
Let $range_{l}^{+}(v)$ be an over- and
$range_{l}^{-}(v)$ an under-approximation of $range_{l}(v)$.
The sets $range_{l}^{-}(v)$ can be realized using a test suite $t \in T_{P}$. All occurring variable values during the tests $\alpha \in t$
can be used as lower bound for $range_{l}(v)$.
On the other hand $range_{l}^{+}(v)$ can be realized using a conservative data flow analysis.
This usually results in
additional
variable values, which never can be achieved.
Both $\delta_{\gamma}^{-}$ and $\delta_{\gamma}^{+}$ can be defined similar
to $\delta_{\gamma}$ using $range_{l}^{-}(v)$ and $range_{l}^{+}(v)$ instead of $range_{l}(v)$.
It is easy to realize, how to bound the gap $\delta_{\gamma}$ using $range_{l}^{-}(v)$ and $range_{l}^{+}(v)$:
\begin{lemma}
    $\delta_{\gamma}^{+} \leq \delta_{\gamma} \leq \delta_{\gamma}^{-}$.
\end{lemma}
Obviously the gap approximation is perfect and an exact gap is found, if $\delta_{\gamma}^{+} = \delta_{\gamma}^{-}$.
In this case it is not necessary to compute exact variable ranges.

\vspace{-0.8em}
\section{Exemplary Illustration}

As a comparative measurement of our method the values
calculated by gcov \cite{gcov} \footnote{\url{http://gnu.org/software/gcov}} are presented at the
end of this section.
The free tool gcov calculates the code coverage during an execution,
which can be used to track the code coverage of a test suite.

\begin{figure}
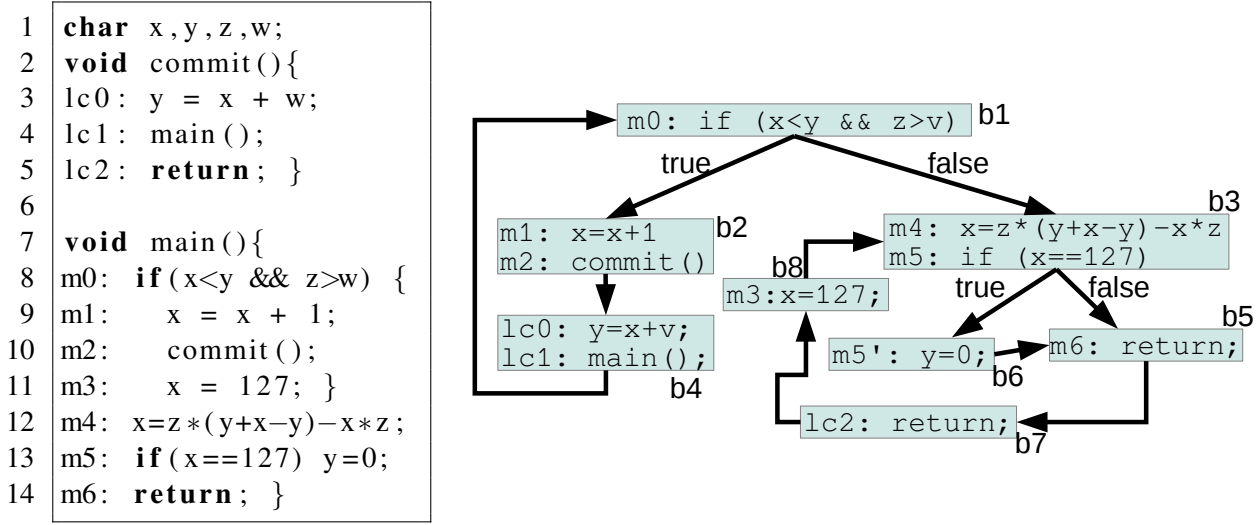

\begin{minipage}[t!]{0.3\textwidth}
\begin{lstlisting}[frame=single,numbers=left]
char x,y,z,w;
void commit(){
lc0: y = x + w;
lc1: main();
lc2: return; }

void main(){
m0: if(x<y && z>w) {
m1:   x = x + 1;
m2:   commit();
m3:   x = 127; }
m4: x=z*(y+x-y)-x*z;
m5: if(x==127) y=0;
m6: return; }
\end{lstlisting}
\end{minipage}~
\begin{minipage}[t!]{0.7\textwidth}
\pgfdeclareimage[interpolate=true,width=11cm]{pic2}{cfglisting2}
\pgfuseimage{pic2}
\end{minipage}
\caption{SPDS example $P_{2}$ in ISO-C syntax for dead code by bad condition + corresponding BBICFG}
\vspace{-0.2cm}
\label{example:dead}
\end{figure}

To show 
the concepts presented so far, we use example $P_{1}$ in Fig. \ref{bsp} and example $P_{2}$ in Fig \ref{example:dead}. 
The constructs in the presented ISO-C code are automatically mapped to SPDS-statements as described in section \ref{isomodel}.
$P_{1}$ contains an arithmetic exception, caused by a
division-by-zero. Hence $P_{1}$ contains a lot of dead code and any test suite with at least one
test would be complete (i.e. there is no way to cover more code).
There is no test necessary ($t=\emptyset$), because
the variables $x$ and $y$ are initialized on labels $l0$ and $l1$. 
Thus it is $range_{l}(x)=range_{l}(x)^{-}=\{0\}$ and $range_{l}(y)=range_{l}(y)^{-}=\{1\}$ for all $l \in L$, where $L=\{l0,l1,l2,lb0,lb1\}$.
It is $range_{l}(x)=range_{l}(x)^{-}=range_{l}(y)=range_{l}(y)^{-}=\emptyset$ for all $l \in labels(P_{1}) \setminus L$.
All the conditions in conditional branches are considered to be statements (see BBICFG in the right part of Fig. \ref{bsp}),
because a conditional branch can contain a statement (e.g. $\texttt{x=0}$ in "$\texttt{if~(x=0)...}$").

Thus it is $|stats(P_{1})| = 15$, $|blocks(P_{1})| = 12$,
$|edges(P_{1})| = 15$, $BExpr(P_{1})=\{ (lc0,x==0), (lb0,y<x) \}$ and $exval(t,P_{1})=\{((lb0,y<x),false)\}$.
Additionally $P_{1}$ is supplemented with variables $v,v_{in}, v_{x==0}$ and $v_{y<x}$ to program $P_{1}'$, where
the $range$ values can be computed accordingly. An inter-procedural conservative interval analysis \cite{muchnick,polyrange} can detect
$range_{l}(v)=range_{l}(v_{in})=\emptyset$
for all $l \in L'=\{lb0', lc2\}$ and $range_{l}(v)=\{0,1\}$ for all $l \in labels(P_{1}') \setminus L'$.
This is used to compute $\delta_{f}^{+}$, $\delta_{b}^{+}$ and $\delta_{s}^{+}$.
The edges $(b2,b3),(b7,b9),(b9,b2),(b3,b10),(b3,b5) \in edges(P_{1})$ are never executed, which is
discovered by the interval analysis. This
results in $\delta_{d}^{+}(P_{1})=\frac{5}{15}$.
Thus, the coverage metrics and gaps of Table \ref{table:gaps1} 
can be calculated for $P_{1}$ as described in 
the previous sections. 
The values were obtained using our current implementation of 
the program described in \cite{kps09dirk}. Computing the values presented in Table
\ref{table:gaps1} takes less than two seconds on a modern Core i7 CPU equipped 
with 8 GiB RAM. 
Table \ref{table:gaps1} also contains
the approximated values as presented in section \ref{sect:approx}.
Although the coverage metrics are far
less than 100 \%, the test suite $t$ is complete.
Additional tests can not improve these coverages as confirmed by the gaps.


\begin{table}[h]
\begin{center}
\begin{tabular}{|c|c|c|c|c|c|c|c|c|c|c|c|c|c|c|c|c|c|c|c|c|c|c|c|c|} \hline
  & $\gamma_{f}(t)$ & $\delta_{f}^+ $& $\delta_{f}$ & $\delta_{f}^-$ & $\gamma_{s}(t)$ & $\delta_{s}^+$ & $\delta_{s}$ & $\delta_{s}^-$ & $\gamma_{d}(t)$ & $\delta_{d}^{+}$ & $\delta_{d}$ & $\delta_{d}^{-}$  \\ \hline
$P_{1}$  & 0.5 & 0.0 & 0.5 & 0.5 & 0.33 & 0.13 & 0.67 & 0.67 & 0.13 & 0.33 & 0.87 & 0.87 \\ \hline
$P_{2}$  & 0.5 & 0.0 & 0.0 & 0.5 & 0.36 & 0.0 & 0.09  & 0.64 & 0.20 & 0.0 & 0.20  & 0.80 \\
\hline \hline
 & $\gamma_{b}(t)$ & $\delta_{b}^{+}$&$\delta_{b}$ & $\delta_{b}^{-}$ & $\gamma_{c}(t)$ & $\delta_{c}^{+}$&$\delta_{c}$ & $\delta_{c}^{-}$ & & & & \\ \hline 
$P_{1}$  & 0.25 & 0.17 & 0.75 & 0.75 & 0.25 & 0.50 & 0.75 & 0.75 & & &  & \\ \hline 
$P_{2}$  & 0.38 & 0.0 & 0.13  & 0.62 & 0.75 & 0.0 & 0.13 & 0.25 & & & & \\ \hline 
\end{tabular}
\end{center}
\vspace{-0.3cm}
\caption{Code Coverages and Gaps for $P_{1}$ in Fig. \ref{bsp} and $P_{2}$ in Fig. \ref{example:dead}}
\label{table:gaps1}
\end{table}


\vspace{-0.5cm}
\begin{table}[h]
\begin{center}
\begin{tabular}{|c|c|c|c|c|c|c|c|c|c|c|c|c|c|c|c|c|c|c|c|c|c|c|c|c|} \hline
$label(P_{2})$ & $range_{*}^{-}(x)$ & $range_{*}^{-}(y)$ & $range_{*}^{-}(z)$ & $range_{*}^{-}(w)$ \\ \hline 
m1,m2,m3,m5',lc0,lc1,lc2 &  $\emptyset$ & $\emptyset$ & $\emptyset$ & $\emptyset$ \\ \hline{}
m0,m4 &  $\{0,1,10\}$ & $\{1,5\}$ & $\{0,1,10\}$ & $\{1,5\}$ \\ \hline{}
m5,m6 &  $\{0\}$ & $\{1,5\}$ & $\{0,1,10\}$ & $\{1,5\}$ \\ \hline
\end{tabular}
\end{center}
\vspace{-0.3cm}
\caption{$range_{l}^{-}$ in $P_{2}$ using test suite $t$ for $P_{2}$ in Fig. \ref{example:dead}}
\label{table:ranges}
\end{table}
\vspace{-0.2cm}

The test suite $t$ discovers $range_{l}^{-}(v)=\{0,1\}$ for each reachable label $l$ in $t$.

Most compilers, i.e. GCC and CL\footnote{Shipped with Microsoft Visual Studio}
from Microsoft, are not able to do a flow-sensitive,
context-sensitive inter-procedural analysis needed for a more precise lower bound in this example.
The abstract interpretations done in a compiler or analysis tool do not yield such a precise lower bound,
as most other tools are essentially model-checkers.
Hence, the lower bound on the range for \texttt{x} and \texttt{y} would include all possible values
at label \texttt{lb1} in $P_{1}$. 
Thus the lower bound on the function gap would be $0$.

Additionally, using the tool gcov to compute the coverage of the test suite of example $P_1$, no
coverage is achieved by any test suite, because gcov does not take arithmetic exceptions into
account resulting in $0\%$ coverage. Of more practical relevance is the calculation of the coverage
gap for non-arithmetic errors. For instance example $P_{2}$ in Fig. \ref{example:dead} has a
difficult condition $(x < y ~\&\&~ z > w)$.
%
%
The variables \texttt{x}, \texttt{y}, \texttt{z} and \texttt{v} of $P_{2}$ (Fig. \ref{example:dead})
are global variables of type $char=bits(..)=8$.
The whole block below ($m1-m3$) becomes dead, if the condition on label $m0$ evaluates to $false$.
Thus \texttt{commit} is not called and the (indirect) recursion not started.
Additionally, for all possible test cases, the
condition ($x == 127$) on label $m5$ never evaluates to $true$.

Let $t = \{(0,1,0,1), (1,1,1,1),(10,5,10,5)\}$
be a test suite with $(x,y,z,w)$ being the values set before calling \texttt{main}.
It is $|stats(P_{2})| = 11$, $|blocks(P_{2})| = 8$,
$|edges(P_{2})| = 10$, $BExpr(P_{2})=\{ (m0,x<y), (m0,z>w), (m0,x<y \&\& z>w), (m5,x==127) \}$ and
$exval(t,P_{2})=\{ ((m0,x<y),true), ((m0,x<y),false), ((m0,z>w),true), ((m0,z>w),false), ((m0,x<y \&\& z>w),false), ((m5,x==127),false) \}$.
The test $\alpha = (0,1,1,0)$ would be a good candidate for the test suite $t$, because $\gamma_{s}(\{\alpha\})=91\%$ is perfect
(proofed by the gap $\delta_{s}$).
Table \ref{table:gaps1} contains 
the calculated coverage metrics and gaps for $P_{2}$.
%
As one can see from the third line of Table \ref{table:gaps1}, the exact gap in the existing code is
rather small: it consists of the condition $x == 127$ on label $m5$ and the following code block.
This is one of the examples in which our method can instruct the tester to expand the test suite.
More code can not be covered, because in each test the variables $x$ and $z$ as well as $y$ and $v$ are aliases.
Although most of the code in the example is alive.
As seen in the previous example, the approximated lower and upper bounds are not perfect. 
An upper
bound on the gap $\delta_{f}$ of called functions, is $\delta_{f}^- = 1 - 0.5$, whereas from the two
available functions one was called during the execution of test suite $t$. 
It is $range_{l}(v)=\{0,1\}$ for every label $l \in labels(P_{2}')$ in $P_{2}'$ (supplementation of $P_{s}$).
$P_{2}$ is also supplemented with variables $v_{in}, v_{x==0}$ and $v_{y<x}$,
so that
the $range$ values can be computed and approximated using an inter-procedural conservative interval analysis ($range^{+}$).
The results for the variables are shown in Table \ref{table:ranges} and \ref{table:gaps1}.

Contrary to gcov the computation of the code coverages followed the C Program and did not rely on
any symbolic assembler. Such abstractions might cause more coverage shown than the actual coverage
in ISO-C. The statement coverage reported by gcov corresponds to $\gamma_s$. Values close or
exactly corresponding to $\gamma_X$ can be obtained from gcov for these particular examples.
Not all values will match $\gamma_X${}, because gcov uses a
different definition for decision and branch coverage and relies on symbolic
assembler output.

\vspace{-0.8em}
\section{Related Work}
A lot of research is done
to get a better coverage for a test suite \cite{enhancing}.
However an important point is often missing:
often it is
impossible to cover 100\% of the code in practice, because of gaps. 

To the best of our knowledge no research has been
done to compute 
provable exact gaps 
used in code coverage.
Conservative strategies 
underestimate the coverage gap \cite{framac}.
Current research only approximates
the gap. 
For instance \cite{structcover} presents a method to automatically add tests by
computing a gap of code covered by the test suite and possible code coverage. 
The authors miss the important point of having code which will and can never be used.
In \cite{structcover} the emphasis is on large scale projects, but especially in such
large projects there is code which cannot be executed and should be removed by the compiler.
As \cite{gittens} describes, some code parts are more important than others. Testing
parts of a program which will never be executed is then a loss of resources. 
Gittens et al. use a domain expert to categorize the code, i.e. for which parts of the
source code their tool should generate tests automatically.  Our gap computation
presented in this paper
could be used
to automatically categorize the code and not depend on a domain expert.
Another project
of interest is \cite{graybox} by Kicillof et al., which shows how to create checkable
models. The focus of Kicillof et al. are models which can be created by stakeholders or maybe even
marketing experts,
and thus is directed at their specific problems at Microsoft. Most other research concerning
the computation of gaps in coverage targets the pre-silicon design validation, i.e. \cite{tempspec, revisited}.

Both papers on pre-silicon design validation are not concerned with testing gaps.
They rather check if a
specification can be achieved.
However, our paper is concerned with languages similar to C, not
any Register Transfer Language (RTL) or even specifications. 

As Regehr correctly writes in \cite{regehr}, such specifications, which are checked in
\cite{tempspec, revisited} might have been wrong in the first place. One solution
proposed by Regehr for finding errors in specifications is having more people to look over these.
A different
solution uses our tool and computes parts of the realized specification that are never used, thus
giving hints to erroneous specifications.

Whereas Berner et al. are targeting the user of an automatic test system \cite{enhancing} our method
targets the automatic test system itself. Berner et al. describe \textit{lessons learned} from their
experience with code coverage analysis tools and automatic test generation tools and propose a list of
rules to be followed when introducing and using an automatic test tool. Our research was not concerned
with usability and group dynamics in a programming environment.

To the best of our knowledge the current research in testing, be it concolic\footnote{interwoven concrete
and symbolic execution} or
model-based, is not concerned with the actual problems of code coverage gaps.
Gap coverage analysis is
not only useful in test case generation but also in the verification of functional
correctness. Imagine the case of a dead function granting more user rights, it is easy to use
a buffer overflow to trigger this functionality. Similar methods have been used by the CCC
for analyzing and using a trojan horse \cite{ccc}\footnote{especially the
section \textit{Upload- und Execute-Mechanismus}}.

Another important tool, which might be able to compete with our method is Frama-C \cite{framac}\footnote{\url{http://frama-c.com}}.
Frama-C is a conservative 
analysis tool, which is
able to find dead code, execute a static value analysis and, contrary to gcov, is able to detect
runtime-errors triggered for instance by division-by-zero. One of the differences between Frama-C and
the method we propose in this paper is the theory behind it.
In contrast to Frama-C \cite{framac}
our method uses exact computation, does not overapproximate the values and does not rely on an
experienced user.
Our exact value analysis produces neither false negatives nor
false positives as in Frama-C.
Although their value analysis sometimes detects that a function does not terminate,
it cannot be used to prove that a function terminates in general.

Frama-C provides sophisticated plugins, but not all of them handle recursion properly.
No sophisticated examples can be handled by Frama-C's value analysis.
Some of the examples tested even cause runtime-errors in Frama-C itself, thus it is
not reliable\footnote{It should be noted, that these runtime-errors should vanish in future versions}.

As our review of the research indicates, none of current research done in testing is
concerned with
\emph{exact} gap computation.

\vspace{-0.8em}
\section{Summary and Conclusions}




This paper presents a framework to compute exact gaps between the feasible and theoretical maximal possible code coverage value.
For specifying programs in an ISO-C semantic we use a very powerful model, namely SPDS. The power of
SPDS 
allows to model an ISO-C compatible semantics for programs without 
abstraction. Therefore we are able to do an
exact value analysis using model checking techniques and so we obtain exact gaps.
We describe how to efficiently approximate the gap in all the other cases.
When using flow-sensitive, path-sensitive, inter-procedural and context-sensitive data flow analyses for
approximating the exact values one can also
use a model-checking tool. The biggest problems of using a model-checker 
are 
false positives or false negatives caused by abstraction. Instead, our approach does not rely on such heavy
abstraction and does not cause false alarms on our ISO-C compatible semantic.
Thus user input or feedback is not required to decide about false alarms.
A lot of computing power is required for using such powerful models. Due to 
smaller programs and smaller data types our approach is still practical 
for embedded systems.

Having combined the best parts of model-checking and static analyses we use expansive model-checking only when
needed (e.g. the gap approximation bounds are not small enough).
Thus the computation of $Post^{*}$ is
needed only if the gap approximation using static analysis and a test suite is not exact ($\delta^{-}\neq \delta^{+}$).

Using our method a lot of metrics can be compared better among
each other now, because of 
exactly specified gaps.
%
Our method allows the testing of non-functional requirements, too. For example the worst case execution times (WCET) using a WCET
metric\footnote{e.g. $\gamma_{WCET}(t):=\max_{a \in t} runtime(a)$ with supplemetation $tick=tick+1$ on each statement,
such that $\gamma_{WCET}(t)=\max (\cup_{l \in labels(S)} range_{l}(tick))$.}
can be computed.

 Our current research considers the practical relevance of exact gap computation for verification of software
 especially in the area of compiler correctness. Additionally we are considering other
 areas of research to apply 
 the computation of exact values and exact gaps. 
 For example the computation of exact value ranges can be used for verification of components \cite{dirkanbo}.



\vspace{-0.8em}
\bibliographystyle{eptcs}
\bibliography{literatur}
\end{document}